\documentclass[journal,a4,draftclsnofoot]{IEEEtran}
\onecolumn
\usepackage{graphicx,amsmath,amssymb,amsfonts,mathtools,adjustbox,arydshln}
\usepackage{times,balance,cite,pgf,tikz,pgffor}
\usepackage[latin1]{inputenc}
\usepackage{mathabx}
\usepackage{setspace}
\usepackage{psfrag}
\usepackage{cite}
\usepackage{xspace}
\usepackage{multirow}
\usepackage{wasysym}
\usepackage{pgfplots}
\usepackage{hyperref}

\newtheorem{theorem}{Theorem}

\newtheorem{lemma}[theorem]{Lemma}
\newtheorem{example}{Example}

\newcommand{\C}{\mathcal C}

\newcommand{\wt}{\textrm{wt}_\textrm{H}}

\newcommand{\sed}[1]{|| #1 ||^2}

\usepackage{stmaryrd}

\usepackage{etextools}
\usepackage{pgf,pgffor}
\makeatletter
\def\and{&\xspace}
\def\rvecpmatrix#1{%
 \begin{pmatrix}%
 \@for\next:=#1\do{\next \and \ \ \ } \!\!\!\!\!\! 
\end{pmatrix} }
\makeatother

\def\rveccommas#1{ \big[ \foreach \x  [count=\ni] in {#1}  {%
\ifnum\ni=1%
\ \x
\else%
 , \x%
\fi%
} \ \big] }

\newlength{\outerwidth}
\newlength{\tbmargin}
\newlength{\bottommargin}

\newcommand{\keynote}[4]{
\setlength{\outerwidth}{#1}
\setlength{\tbmargin}{#2}
\setlength{\topmargin}{\tbmargin}
\setlength{\bottommargin}{\tbmargin}
\advance\topmargin by -24pt
\colorbox{white}{%
\fbox{\begin{minipage}{\outerwidth}
{\color{white} \rule{\linewidth}{0.4mm}\\[\topmargin]}
\begin{center}
\makebox[0pt][c]{
\advance\outerwidth by -#3
\begin{minipage}{\outerwidth}
\setlength\parindent{0cm}
\setlength\parskip{12pt}
#4
\vspace{\bottommargin}
\end{minipage}
}\end{center}%
\end{minipage}}
}
}

\renewcommand{\v}{{\sigma^2}}

\renewcommand{\cal}{\mathcal}

\renewcommand{\v}{\mathbf}

\newcommand{\tr}{^\mathrm t}

\newcommand{\bbr}{\mathbb R}
\newcommand{\bbrn}{\mathbb R^n}
\newcommand{\bbz}{\mathbb Z}
\newcommand{\bbzn}{\mathbb Z^n}

\newcommand{\x}{\v x}
\newcommand{\y}{\v y}
\newcommand{\z}{\v z}

\def\slantfractiny#1#2{\tiny {\hbox{$\,^#1\!/_#2$}}}
\newcommand{\oh}{{\slantfractiny{1}{2}}}
\newcommand{\oq}{{\slantfractiny{1}{4}}}

\renewcommand{\b}[1]{\mathbf{#1}}
\renewcommand{\L}{\Lambda}
\newcommand{\Lc}{\Lambda_{\mathrm c}}
\newcommand{\Ls}{\Lambda_{\mathrm s}}
\newcommand{\G}{\mathbf G}
\renewcommand{\H}{\mathbf H}
\newcommand{\Gc}{\mathbf{G}_{\mathrm c}}
\newcommand{\Gs}{\mathbf{G}_{\mathrm s}}
\newcommand{\Hc}{\mathbf{H}_{\mathrm c}}
\newcommand{\Hs}{\mathbf{H}_{\mathrm s}}

\newcommand{\V}{\mathcal V}

\renewcommand{\index}{\mathrm{index}}
\newcommand{\enc}{\mathrm{enc}}

\newcommand{\iif}{\textrm{ if }}

\renewcommand{\and}{\textrm{ and }}
\newcommand{\aand}{\textrm{ and }}

\newcommand{\oor}{\textrm{ or }}
\newcommand{\otherwise}{\textrm{ otherwise }}

\newcommand{\figref}[1]{Fig.~\ref{#1}}

\renewcommand{\cal}{\mathcal}

\usepackage{bm}

\renewcommand{\P}{\mathsf{P}}

\newcommand{\W}{\mathsf{W}}

\definecolor{wesBlue}{HTML}{1DADE8}
\definecolor{wesRed}{HTML}{F34D29}

\newcommand{\diag}{\mathrm{diag}}
\newcommand{\ri}{rectangular encoding\xspace}
\newcommand{\ptope}{parallelotope\xspace}
\newcommand{\ptopes}{parallelotopes\xspace}
\newcommand{\M}{\b M}
\newcommand{\I}{\b I}
\renewcommand{\P}{\b P}
\renewcommand{\H}{\b H}
\renewcommand{\W}{\b W}
\renewcommand{\wt}[1]{\widetilde{\b #1}}

\IEEEoverridecommandlockouts   

\author{Brian M.~Kurkoski\thanks{Author contact: kurkoski@jaist.ac.jp. This paper was presented in part at the Information Theory and Applications Workshop \cite{Kurkoski-itw15}, the Symposium on Information Theory and Its Applications \cite{Kurkoski-sita15} and the 9th Asian-European Workshop on Information Theory \cite{Kurkoski-aew9}. }}

\begin{document}
\title{Encoding and Indexing of Lattice Codes}

\maketitle

\begin{abstract}
Encoding and indexing of lattice codes is generalized from self-similar lattice codes to a broader class of lattices.  If coding lattice $\Lc$ and shaping lattice $\Ls$ satisfy $\Ls \subseteq \Lc$, then $\Lc / \Ls$ is a quotient group that can be used to form a (nested) lattice code $\C$.  Conway and Sloane's method of encoding and indexing does not apply when the lattices are not self-similar.  Results are provided for two classes of lattices. (1) If $\Lc$ and $\Ls$ both have generator matrices in triangular form, then encoding is always possible. (2) When $\Lc$ and $\Ls$ are described by full generator matrices, if a solution to a linear diophantine equation exists, then encoding is possible. In addition, special cases where $\C$ is a cyclic code are also considered.  A condition for the existence of a group homomorphism between the information and $\C$ is given. The results are applicable to a variety of coding lattices, including Construction A, Construction D and LDLCs.  The $D_4$, $E_8$ and convolutional code lattices are shown to be good choices for the shaping lattice.  Thus, a lattice code $\C$ can be designed by selecting $\Lc$ and $\Ls$ separately, avoiding competing design requirements of self-similar lattice codes.
\end{abstract}

\section{Introduction}

\subsection{Motivation}

An $n$-dimensional lattice $\L$ is an additive subgroup of $\bbrn$.   Let $\Lc$ and $\Ls$ be two lattices such that $\Ls \subseteq \Lc$.  Then, $\Lc / \Ls$ forms a quotient group.  If the coset leaders $\cal C$ of this quotient group are chosen from the zero-centered Voronoi region of $\Ls$, this construction is called a nested lattice code.  Nested lattice codes are well-suited  for channels such as wireless communications: the coding lattice $\Lc$ provides coding gain; the shaping lattice $\Ls$ provides shaping gain; and group properties make $\C$ a candidate for physical-layer network coding.

Encoding is mapping information integers to the codewords of $\C$.  Indexing is the inverse operation, mapping codewords of $\C$ to information integers.  For self-similar lattices, $\Ls = K \Lc$ with $K \in \mathbb Z$, Conway and Sloane gave an efficient algorithm to perform encoding and indexing \cite{Conway-it83}.

However, self-similar lattices are less suitable for practical implementations, because of competing design requirements for $\Lc$ and $\Ls$.  In particular, high coding gain lattices $\Lc$ are typically decoded using belief propagation algorithms which allow the lattice dimension $n$ to be high. On the other hand, high shaping gain lattices should possess an efficient quantization algorithm needed for the modulo shaping operation.  But these belief-propagation decoded coding lattices are poor choices as a shaping lattice, due to the complexity of the quantization algorithm; on the other hand, lattices with good shaping gain and efficient quantizaiton algorithms do exist.   Thus, it is desirable to select $\Lc$ and $\Ls$ to be not self-similar. As long as $\Ls$ is a sublattice of $\Lc$, the quotient group $\Lc/\Ls$ exists, and the nested lattice code\footnote{Because $\Ls \subseteq \Lc$ is assumed, all lattice codes in this paper are nested so this is last mention of ``nested lattice codes'' --- from here $\C$ is referred to as a lattice code. Nested binary codes are later used for Construction D lattices.} $\cal C$ can be constructed.    However, in this generalized scenario when $\Ls$ and $\Lc$ are not self-similar, an unexpected problem arises: Conway and Sloane's encoding and indexing cannot be used --- that is, even though the quotient group exists, it is not clear how to map information to the codebook $\C$.

\subsection{Contributions}

This paper generalizes encoding and indexing of lattice codes based on self-similar lattices $\Ls = K \Lc$ to a broader class of lattices $\Lc$ and $\Ls$.  A \emph{\ri} is defined, where each information integer $b_i$ is from the set $\{0,1, \ldots, M_i -1\}$, for $i=1,2, \ldots, n$, that is information integers can be selected independently in each dimension.  In a \ri, the information vector $\b b = [b_1, \ldots, b_n]\tr$ is systematically and bijectively encoded to the elements of $\C$.  An observation is that if $\Lc$ has a matrix of basis vectors $\Gc$ that is ``aligned'' with $\Ls$, then this basis can be used for rectangular encoding.  Stating this condition technically, if the $n$ basis vectors $\Gc$, when scaled by $M_1, \ldots, M_n$ respectively, form a fundamental \ptope of $\Ls$, then this basis $\Gc$ and the $M_i$ should be used for encoding.  Preliminaries on lattices, lemmas that describe \ptope fundamental regions, and lattice cosets, are given in Section \ref{sec:preliminaries}. The rectangular encoding and this technical condition are given in Section \ref{sec:rectangularEncoding}.

This paper gives two cases where \ri is possible.  In the case where the bases of $\Lc$ and $\Ls$ can both be written as triangular matrices, this technical condition is always satisfied, and a rectangular encoding exists.  This is an effective way to encode lattice codes of high dimension; it is described in Section \ref{sec:triangular}.  In the more general case of full matrices, the technical condition may not be satisfied for the given basis $\Gc$, however it may be possible to find an alternative basis which does satisfy this condition. This basis transformation is possible if a linear diophantine equation derived from the generator matrices has a solution.  In addition, special cases where $\C$ is a cyclic code are also considered. The full-matrix lattice case is described in Section \ref{sec:full}.

The information integers $b_i \in \bbz_{M_i}$ for $i=1, \ldots, n$ are regarded as elements of the group $\bbz / {M_i} \bbz$.  A homomorphism between the information vector $\b b$ and the lattice code $\C$ is potentially useful for lattice-based physical-layer network coding.  A condition on the lattice generator matrices is given in Section \ref{sec:homo}; if satisfied, then a group homomorphism exists.  The paper concludes with discussion in Section \ref{sec:discussion}.

Of interest are lattice codes which have shaping gain provided by the Voronoi region of $\Ls$. Shaping requires a modulo-lattice operation, and the modulo lattice operation requires quantization in $\Ls$; this paper writes ``quantization'' when referring to the need to construct a shaped lattice code $\C$.   There is no known polynomial-time algorithm for optimal quantization in general, and complexity increases dramatically in the lattice dimension $n$, but there exist lattices with good shaping gain and reasonable quantization complexity.  The motivation of this paper is practical encoding of lattice codes with reasonable shaping gain, reasonable quantization complexity, and excellent coding gain.  Because the objective in this paper is encoding and indexing, the error correction ability of the code, which is provided by $\Lc$ does not need to be studied --- in fact the encoding scheme does not change the probability of a lattice error.

The results in this paper are applicable to a wide variety of lattices, so long as the lattice generator matrices, or their inverse, are known.   Throughout this paper low-dimensional examples are used.  The example coding lattices include those formed using Construction A and Construction D \cite{Conway-1999} and low-density lattice codes (LDLC) \cite{Sommer-it08}. The examples of shaping lattices include and $D_n$, $E_8$ and convolutional code lattices.  The examples both illustrate the principles of encoding, and show the wide range of lattices to which the proposed techniques are applicable. The examples also illustrate how to match the dimension of a shaping lattice to the coding lattice.

\subsection{Related Work}

The ideal shaping region is an $n$-dimensional sphere, which has a maximum possible shaping gain of 1.53 dB as $n\to \infty$ \cite[Ch.~14]{Forney-2005}.  Hyperspherical shaping regions are impractical except in small dimension, so it is fortunate that the Voronoi region of many lattices is sphere-like. In the domain of trellis codes for the AWGN channel, Forney showed that convolutional codes can be used for shaping, showing how to obtain much of the maximum possible shaping gain \cite{Forney-it92}. Erez and ten Brink used such shaping to design a close-to-capacity coding scheme, for the known-interference channel \cite{Erez-it05*3}.  There is however a lattice formulation of trellis codes \cite[Ch.~3]{Schlegel-1997}, and it is the lattice-theoretic model of coding for the AWGN channel that has received attention, recently.

Numerous new constructions for coding lattices have appeared in the literature.  These are high-dimension lattices, decoded using belief propagation, that offer high coding gain, often close to the Poltyrev limit \cite{Poltyrev-it94}, but are unconstrained lattices.  The following such lattices are relevant to the examples in this paper.  Yan et al.~formed polar lattices using Construction D, and a finite-length code comes within 1.6 dB of the Poltyrev capacity \cite[Sec.~3.5]{Yan-2014} (see also \cite{Yan-isit13}).  Construction D is important because it uses binary codes, which are widely understood; such lattices have also been formed from LDPC codes \cite{Sadeghi-it06} and turbo codes \cite{Sakzad-arxiv11} \cite{Sakzad-aller10}.  Non-binary LDPC code lattices, called LDA lattices, were introduced by di Pietro, Boutros, Z{\'e}mor and Brunel, are formed using using Construction A, and come within 0.7 dB of the Poltyrev limit for a finite-length code \cite{diPietro-itw12}.  Spatially-coupled LDA lattices come within 0.2 dB of the Poltyrev limit \cite{Tunali-itw13}.  Sommer, Shalvi and Feder proposed LDLC lattices, which explicitly construct a sparse lattice check matrix; belief-propagation decoding within 0.6 dB of the Poltyrev limit was claimed \cite{Sommer-it08}.

Some of the above lattices have been shaped using self-similar lattices. For LDLC lattices, Sommer et al.~showed a 0.4 dB shaping gain, using the M-algorithm to perform the quantization operation \cite{Sommer-itw09} (see \cite{Ferdinand-asilomar14} for further validation), and a similar 0.4 dB gain was observed when using a belief-propagation algorithm for the quantization operation \cite{Kurkoski-isit09*2}.  For LDPC code lattices, Khodaiemehr, Sadeghi and Sakzad showed a 0.63 dB shaping gain using integer least-squares optimization \cite{Khodaiemehr-arxiv16}.  These results illustrate the problem of self-similar lattices: the quantization algorithms (i.e.~modulo lattice operation) are computationally complex, and yield relatively modest shaping gains.  Note that hypercube shaping is computationally trivial, when the lattice matrix has a triangular form \cite{Sommer-itw09}, or if the lattice is based on Construction A or Construction D, but this offers no shaping gain. 

On the other hand, 0.65 dB shaping gain can be obtained with low complexity, and up to 1.36 dB shaping gain with modest complexity has been claimed.  The $E_8$ quantization algorithm is very simple, and provides 0.65 dB of shaping gain. The $n=16$ Barnes-Wall lattice has 0.86 dB shaping gain, and the $n=24$ Leech lattice has 1.03 dB shaping gain, and their quantization algorithms have complexity low enough to be practical.  Another major approach to shaping is to use convolutional codes, where the Viterbi algorithm implements quantization.  The shaping gain increases with the number of trellis states, as much as 1.36 dB was claimed possible by Forney \cite{Forney-it92}, citing \cite{Marcellin-com90}, in the context of coded modulation. 

One success in the direction of shaping high dimension lattices is by Sommer et al., who described a systematic technique to encode integers to LDLC lattice points \cite{Sommer-itw09}. If the integers are pre-shaped with the $E_8$ or other lattice, then the resulting code has much of that lattice's shaping gain \cite{Ferdinand-itw14}. This is an effective technique to shape LDLC lattices, but there is a  penalty at low rates, and the sublattice condition is not satisfied, that is the quotient group $\Lc / \Ls$ does not exist.

The contribution of this paper on encoding non-self-similar lattice codes is distinct from past work.  The encoding methods are  different from trellis coding, since the underlying structure is a coding lattice $\Lc$,  and the elegant structure of the coding lattice is preserved. Shaping self-similar lattices of high dimension is computationally difficult, and only yields moderate benefits.  On the other hand, this paper shows that it is possible to select a shaping lattice which is not self-similar, while still possessing many desirable properties.

\section{Preliminaries}

\label{sec:preliminaries}

The following notation is used.  Bold face uppercase letters $\G$ denote matrices, and the components are lowercase of the same letter $g_{i,j}$ when possible; $\I_n$ is the $n$-by-$n$ identity matrix. Bold face lowercase $\x$ are column vectors, and the components are lowercase of the same letter $x_i$, and $[\ ]\tr$ denotes transpose, so $\x = [x_1, x_2, \ldots, x_n]\tr$.  The index $i$ is usually used so that $i = 1,2, \ldots, n$.  Calligraphic $\cal F$ font denotes a set. The set of real numbers is $\bbr$ and the set of integers is $\bbz$.

\subsection{Lattice Matrix Definition}

A possible basis for $n$-dimensional lattice $\L$ is an $n\times n$ generator matrix $\G$ of full rank.  The corresponding check matrix is $\H = \G^{-1}$, and $\x \in \L$ if and only if $\H \x$ is an integer.  (The rows of $\H$ generate the dual lattice, but in this paper the check matrix interpretation is preferred.)

Two lattices are used, a coding lattice $\Lc$ and a shaping lattice $\Ls$.  The coding lattice has a generator matrix $\Gc$ consisting of generator vectors $\v v_i$ in columns:
\begin{align}
\Gc = 
\begin{bmatrix}
\v v_1 & \v v_1 & \cdots & \v v_n
\end{bmatrix},
\end{align}
and corresponding check matrix $\Hc$.  The shaping lattice $\Ls$ has a generator matrix $\Gs$ consisting of generator vectors $\v g_i$ in columns:
\begin{align}
\Gs = 
\begin{bmatrix}
\v g_1 & \v g_1 & \cdots & \v g_n
\end{bmatrix}
\end{align}
and corresponding check matrix $\Hs$.  Shortest-distance quantization of $\y \in \bbrn$ is:
\begin{align}
Q_{\Ls}(\y) &= \min_{\lambda \in \Ls} \sed{\y - \lambda}. \label{eqn:quantization}
\end{align}

Throughout this paper, it is assumed that $\Ls \subseteq \Ls$, and is referred to as the \emph{sublattice condition}.  Necessary and sufficient conditions for $\Ls \subseteq \Lc$, are well-known \cite[p.~179]{Zamir-2014}, but given here.
\begin{lemma}
$\Ls \subseteq \Lc$ if and only if $\Hc \Gs$ is a matrix of integers.
\end{lemma}

\emph{Proof}   Let $\Gs \b b \in \Ls$.  The point $\Gs \b b$ is a point in $\Lc$ if and only if $\Hc \Gs \b b$ is a vector of integers.   For arbitrary $\b b \in \bbzn$, this is true if and only if $\Hc \Gs$ is a matrix of integers.  \hfill $\blacksquare$

\subsection{Fundamental Region}

This subsection reviews the fundamental region of a lattice, and shows two non-trivial \ptopes which are fundamental regions. A region $\cal F \subset \bbrn$ is called a \emph{fundamental region} for a lattice $\L$ if shifts of $\cal F$ by lattice points covers $\bbrn$ exactly, that is, $\L + \cal F = \bbrn$.  

The volume of a fundamental region $|\cal F|$ is equal to $\det(\G)$, where $\G$ is the generator matrix.  The Voronoi region $\cal V$ for $\L$ is a fundamental region.  Certain \ptopes  are also fundamental regions.  A \ptope $\cal P$ is described by an $n$-by-$n$ full rank matrix $\P$:  
\begin{align}
\cal P(\P) = \big\{ \P \cdot
\begin{bmatrix}
s_1 \\ \vdots \\ s_n
\end{bmatrix}
\mid
0 \leq s_1,s_2, \ldots, s_n < 1
\big\}.
\end{align}
Given a generator matrix $\G$, the natural \ptope $\cal P(\G)$ is a fundamental region.  However, other \ptopes, obtained by modifications to $\G$, may also be fundamental regions.   A \ptope fundamental region satisfies the property that any $\y \in \bbrn$ may be expressed as
\begin{align}
\y = \G \b b + \P \b s 
\end{align}
for a unique integer vector $\b b$ and unique fractional part $\b s$, with $0 \leq s_i < 1$.  

Two \ptope fundamental regions are given in two lemmas.    The following lemma shows that if the lattice generator matrix is triangular, then any triangular matrix $\P$ that agrees on the diagonal elements will form a \ptope fundamental region.

\begin{lemma}
\label{lemma:triangularFundamental} Let $\G$ be a triangular generator matrix for a lattice $\L$.  Let $\P$ be a triangular matrix with the same diagonal elements as $\G$.  Then \ptope $\cal P(\P)$ is a fundamental region for $\L$. 
\end{lemma}

\emph{Proof}  Assume lower triangular matrices, with $\G$ and $\P$ given by:
\begin{align*} 
\begin{bmatrix}
g_{11} & 0 & \cdots & 0 \\
g_{21} & g_{22} & \cdots & 0 \\
\vdots & & \ddots & \vdots \\
g_{n1} & g_{n2} & \cdots & g_{nn}  
\end{bmatrix}
\aand
\begin{bmatrix}
g_{11} & 0 & \cdots & 0 \\
p_{21} & g_{22} & \cdots & 0 \\
\vdots & & \ddots & \vdots \\
p_{n1} & p_{n2} & \cdots & g_{nn}  
\end{bmatrix}
\end{align*}
respectively, where $g_{ij}$ are the given matrix values and $p_{ij}$ are arbitrary values.  Note that $\det(\P) = \det(\G)$ holds by the construction of $\P$. To show $\cal P(\P)$ is a fundamental region, it is sufficient to show that for any $\y \in \bbrn$,
\begin{align}
\y = \G \b b + \P \b s
\end{align}
has a unique solution in $\b b$ and $\b s$, where $b_i$ are integers and $0 \leq s_i < 1$.   For row one:
\begin{align}
y_1 &= g_{11} b_1 + g_{11} s_1
\end{align}
has a unique solution, $b_1$ and $s_1$ are the integer and fractional parts of $\frac{y_1}{g_{11}}$, respectively.  For row two:
\begin{align}
y_2 &= g_{21} b_1 + g_{22} b_2 + p_{21} s_1 + g_{22} s_2
\end{align}
which also has a unique solution, $b_2$ and $s_2$ are the integer and fractional parts of
\begin{align}
\frac{y_2 - g_{21} b_1 - p_{21} s_1}{g_{22}},
\end{align}
respectively.  This continues recursively, so that all $b_i$ and $s_i$ for $i=1,2, \ldots, n$ have unique solutions.  Since there is a unique solution for all $b_i$ and $s_i$, any $\y$ is in exactly one \ptope, and thus $\cal P(\P)$ is a fundamental region. \hfill $\blacksquare$

Recall that $\cal P(\G)$ is a fundamental region. The next lemma shows that if one basis vector of $\G$ is replaced with some other volume-preserving vector (not necessarily a basis vector), that the resulting \ptope is still a fundamental region for the lattice.  Here $\G$ may be a full matrix.

\begin{lemma}
\label{lemma:squareFundamental} Let $\G$ be a basis of $n$ column vectors that generates $\L$. Replace one column vector of $\G$ with any linearly independent column vector to form $\G'$ such that $\det(\G) = \det(\G')$. Then $\cal P(\G')$ is a fundamental region for $\L$. \label{lemma:ptope}
\end{lemma}

\emph{Proof}  To show that \ptope $\cal P(\G')$ is a fundamental region for $\L$, given an arbitrary point $\x \in \bbrn$, it will be shown that:
\begin{align}
\x &= \G \v b + \G' \v s 
\end{align}
has a unique solution, with each $b_i$ an integer and each $0 \leq s_i < 1$.  Multiply both sides by $\G^{-1}$:
\begin{align}
\G^{-1} \x &= \v b + \G^{-1} \G'  \v s. \label{eqn:ptopeproof}
\end{align}
Since $\G$ and $\G'$ differ only in one column $t$, $\G^{-1} \G'$ has the form of an identity matrix with column $t$'s zeros replaced with arbitrary values $\alpha_i$, for example $n=5$ and $t=4$:
\begin{align}
\begin{bmatrix}
1 & 0 & 0 & \alpha_1 & 0 \\
0 &1  & 0 & \alpha_2 & 0 \\
0 &0  & 1 & \alpha_3 & 0 \\
0 &0  & 0 & 1 & 0 \\
0 &0  & 0 & \alpha_5 & 1 \\
\end{bmatrix}.
\end{align}
All diagonal elements are 1 because $\det(\G) = \det(\G')$ was assumed. Let $\y = \G^{-1} \x$.  Row $t$ of \eqref{eqn:ptopeproof} is:
\begin{align}
y_t &= b_t + s_t 
\end{align}
and clearly has a unique solution, $b_t$ and $s_t$ are the integer and fractional parts of $y_t$, respectively.   Any other row $k\neq t$ is:
\begin{align}
y_k &= b_k + s_t \alpha_t + s_k 
\end{align}
which also has a unique solution, $b_k$ and $s_k$ are the integer and fractional parts of $y_k - s_t \alpha_t$, respectively.  Since there is a unique solution for all $b_i$ and $s_i$, any $\y$ is in exactly one \ptope, and thus $\cal P(\G')$ is a fundamental region. \hfill $\blacksquare$

\subsection{Lattice Cosets}

This subsection gives a review of lattice cosets.  See also Zamir's book \cite[Ch.~2]{Zamir-2014}, Forney's review \cite[Sec.~II]{Forney-it88*1}, or references on abstract algebra, e.g. \cite{Gallian-2012}. 

Recall $\Ls \subseteq \Lc$. For any $\x \in \Lc$, the set $\x + \Ls$ is the \emph{coset of $\Ls$ in $\Lc$ containing $\x$}.   Each coset is a set of infinite size.

A \emph{quotient group} $\Ls/\Lc$ is the set of all cosets:
\begin{align}
\Ls/\Lc = \{ \x + \Ls | \x \in \Lc\},
\end{align}
and the number of elements of $\Lc/\Ls$ is finite.  Let $\oplus$ denote addition in the quotient group.  
If $\x, \y \in \Lc$, and $\z = \x + \z$, then $(\x + \Ls) \oplus (\y + \Ls)$ is the coset containing $\z$, that is $\z + \Lc$.  For any integer vector $\b c \in \bbzn$, $\x$ and $\x + \Gs \b c$ are in the same coset.

A \emph{coset leader} is an element of the set $\x + \Ls$ chosen to represent the coset.  A set of coset leaders can be chosen with respect to any fundamental region $\cal F$ of lattice $\Ls$, that is,
\begin{align}
 \Lc \cap \cal F \textrm{ are coset leaders of } \Lc / \Ls.
\end{align}
If $\x \in \Lc$, then the coset leader of $\x + \Ls$ is $(\x + \Ls) \cap \cal F$.  When the fundamental region is the zero-centered Voronoi region $\V$ of $\Ls$, then the set of coset leaders is the codebook, or lattice code $\C$:
\begin{align}
\cal C =  \Lc \cap \V.
\end{align}
This $\C$ is used by the transmitter in a communications system.

Group operations in the quotient group $\Lc/\Ls$ may be performed using coset leaders, since the coset leader represents its coset.   Let  $\x, \y \in \cal C$.  The set $\C$ forms a group under $\oplus$, where $\x \oplus \y = \z$ may be computed as a modulo-$\Ls$ operation:
\begin{align}
\z = \x + \y - Q_{\Ls}(\x + \y),  
\end{align}
using the quantization operation \eqref{eqn:quantization}. 

\begin{lemma}
\label{lemma:twoFundamentalRegions}
With $\Ls \subseteq \Lc$,  let $\cal F_1$ and $\cal F_2$ be two fundamental regions of $\Ls$. Then, there is a bijection between the two sets $\cal C_1$ and $\cal C_2$:
\begin{align}
\cal C_1 = \Lc \cap \cal F_1 \aand \cal C_2 = \Lc \cap \cal F_2.
\end{align}
\end{lemma}

\emph{Proof} Note that $|\cal F_1| = |\cal F_2| = \det(\Gs)$ and that $|\C_1| = |\C_2| = \det(\Gs) / \det (\Gc)$.  The bijective mapping between $\C_1$ and $\C_2$ is through cosets.  An element $\x \in \C_1$ belongs to the coset $\x + \Ls$, and $\x + \Ls \cap \C_2$ consists of exactly one element, namely the coset leader of $\x + \Ls$  in $\C_2$.  \hfill $\blacksquare$

\section{\ri}
\label{sec:rectangularEncoding}

This section defines \ri, gives the key technical lemma, and gives an example that motivates the problem.

\subsection{Rectangular Encoding}

Let $\C$ be a lattice code, given by suitably chosen coset leaders of $\Lc / \Ls$.  The number of codewords $|\C|$ is $M  = | \det ( \Gs) |  / | \det (\Gc)|$, and the code rate is: 
\begin{align}
R = \frac 1 n \log_2 M = \frac 1 n \log_2 \frac{| \det (\Ls) |}{| \det (\Lc)|}. 
\end{align}

\emph{Definition}   The lattice code $\C$ has a \emph{\ri} if there exists $\Gc$ and positive integers $M_1, \ldots, M_n$ 
such that the function:
\begin{align}
\x & = \Gc \b b - Q_{\Ls}(\Gc \b b) \label{eqn:encode}
\end{align}
is a bijective mapping between the integers $b_i \in \{0,1, \ldots, M_i - 1\}$ and the codebook $\x \in \C$.  This encoding operation \eqref{eqn:encode} is abbreviated $\x  = \enc(\b b)$. 

In other words, the encoding generates $\cal C$ exactly.  Let $\M$ be the diagonal matrix with $M_i$ on the diagonal:
\begin{align}
\M &=\diag(M_1,M_2, \ldots, M_n),
\end{align}
with $\det(\M) = M$.  Note $M = \prod_{i=1}^n M_i$ and $\det(\Gc \M) = M \det(\Gc) = \det(\Gs)$.  ``Rectangular'' emphasizes the point that each $b_i$ is selected independently of the other integers; in a less systematic method, the integer range for $b_i$ would depend on integers selected in other positions, which is not desirable. Of course \eqref{eqn:encode} is the standard shaping operation, or lattice modulo operation \cite[p.~21]{Zamir-2014}.

Following Conway and Sloane \cite{Conway-it83}, the inverse operation $\enc^{-1}$ is called indexing, since the vector $\b b$ may be thought of as the index of codeword $\x \in \C$, and is abbreviated:
\begin{align}
\b b &= \index(\x). 
\end{align}
The indexing operation for $\x \in \cal C$ amounts to finding the element of $\x + \Ls$ inside $\cal P(\Gc \M)$.  The method differs for the triangular matrix lattice indexing and full matrix lattice indexing.

\subsection{Key Technical Lemma}

This section gives the key technical lemma that makes a connection between the coding lattice basis $\Gc$ and shaping lattice basis $\Gs$.

The key point is to recognize that encoding with $\Gc$ and a suitable choice of $M_1, M_2, \ldots, M_n$ efficiently labels points of $\Lc$ inside the \ptope $\cal P(\Gc \M)$, that is the points $\Gc \b b$.  If this \ptope is a fundamental region of $\Ls$, then by Lemma \ref{lemma:twoFundamentalRegions}, there is a bijective mapping between the elements of $\Lc \cap \cal P(\Gc \M)$ and $\C$.  Of course $\C = \Lc \cap \V$ and the Voronoi region $\V$ is a fundamental region.  This reasoning proves the following lemma.

\begin{lemma}
\label{lemma:KeyTechnicalLemma}
If $\cal P(\Gc \M)$ is a fundamental region of $\Ls$, then the corresponding $\Gc$ and $\M$ form a \ri for $\cal C$.
\end{lemma}

Self-similar encoding \cite{Conway-it83}  $\Lc/K \Lc$ with $K \in \bbz$ satisfies the condition of Lemma \ref{lemma:KeyTechnicalLemma}. The scaling is $\M = K \I_n$, where $\I_n$ is the identity matrix, so $\Gs = \Gc \M$.  Clearly $\cal P(\Gs)$ is a fundamental region of $\Ls$ and so self-similar encoding forms a \ri by Lemma \ref{lemma:KeyTechnicalLemma}.   More generally, this lemma makes a connection between $\Lc$ and $\Ls$, even if they are not self-similar. If $\Gc$ is ``aligned'' with $\Ls$ as described by the lemma, then a \ri exists.

\subsection{Simple Example of \ri}

The following $n=2$ example illustrates the problem addressed in this paper.  Two lattices that satisfy $\Ls \subseteq \Lc$, are given, but not all choices of $M_1$ and $M_2$ allow for a \ri.

\begin{figure}[t]
\begin{center}
 \input{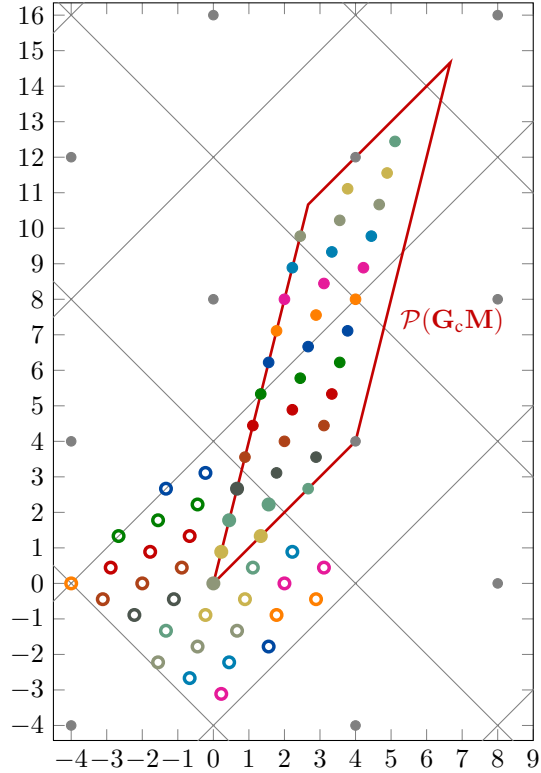} 
\caption{Rectangular encoding for Example 1 using $(M_1,M_2) = (3,12)$.  Gray represents the $4D_2$ shaping lattice.  Dots represent $\Gc \v b$, and the \ptope $\cal P(\Gc \M)$ is shown in red. Open circles are the lattice code $\cal C$ obtained from $\Gc \v b - Q_{\Ls}(\Gc \v b)$.  Colors indicate cosets, although some colors are repeated. \label{fig:example1} }
\end{center}
\end{figure}

\begin{example}
\label{example:basicRi}
Consider $\Ls = 4D_2$:
\begin{align}
\Gs = \begin{bmatrix}
4 & 0 \\ 4 & 8
\end{bmatrix},
\end{align}
which has no shaping gain, but is useful as an example.  Consider $\Lc$ with check matrix:
\begin{align} \Hc =
\begin{bmatrix*}[r]
1 &  - \frac 1 4 \\[2pt] -\frac 3 2 & \frac 3 2
\end{bmatrix*}
\end{align}
which has $\Hc^{-1}$:
\begin{align}
\Gc &=
\begin{bmatrix*}[r]
\frac 4 3 & \frac 2 9 \\[2pt] \frac 4 3 & \frac 8 9
\end{bmatrix*}.
\end{align}
The sublattice condition $\Ls \subseteq \Lc$ is satisfied since $\Hc \Gs$ is a matrix of integers. The number of messages is:
\begin{align}
M &= \det(\Gs) / \det(\Gc)  = 36.
\end{align}

The problem: Is it possible to find $M_1$ and $M_2$ for integer ranges $b_i \in \{0, \ldots, M_i-1\}$ such that integers map bijectively to $\C$, with $M=M_1 M_2$? The number of messages factors as  $2\cdot 2\cdot 3\cdot 3$.  By an exhaustive search on possible factorizations, only the choices $(M_1,M_2) = (3,12)$ and $(1,36)$ yield \ri.  Encoding with $(3,12)$ is shown in \figref{fig:example1}.
 \end{example}

Other choices of $(M_1,M_2)$ do not yield \ri, because two distinct $(b_1,b_2)$ can map to the same element of $\cal C$.  For example, with $(M_1,M_2) = (6,6)$, $\v b = (1,0)$ and $\v b = (4,0)$ both encode $\enc(\v b) = (\frac 4 3, \frac 4 3)$.  At the same, time, half of $\C$, including for example $(- \frac 2 9, -\frac 8 9)$, cannot be encoded.

\newcommand{\bbzstar}{\bbz_*^n}
\newcommand{\cf}{\C_f}

\section{Encoding and Indexing Triangular Matrix Lattices}
\label{sec:triangular}

\subsection{Encoding Triangular Matrix Lattices}

When $\Gs$ and $\Hc$ are both triangular, encoding and indexing are particularly simple; they may be either both upper triangular or both lower triangular. Let $g_{ij}$ and $h_{ij}$ represent the elements of $\Gs$ and $\Hc$ respectively, in row $1 \leq i \leq n$ and column $1 \leq j \leq n$.  

\begin{lemma}
\label{lemma:triangularLattice}
When $\Hc$ and $\Gs$ are triangular with diagonal elements $h_{ii}$ and $g_{ii}$ respectively, \ri can be performed using $\Gc$, and using $M_i$ given by:
\begin{align}
M_i = h_{ii} g_{ii}. 
\end{align}
\end{lemma}

\emph{Proof} Consider the triangular matrix $\P$:
\begin{align}
\P &= \Gc  \M.
\end{align}
The diagonal elements of triangular $\Gc = \Hc^{-1}$ are $\frac 1 {h_{ii}}$, so the diagonal elements of $\P$ are $\frac{M_i}{h_{ii}} = g_{ii}$.  Since the diagonal elements of $\P$ are the same as the diagonal elements of $\Gs$, by Lemma \ref{lemma:triangularFundamental}, $\cal P(\P)$ is a fundamental region for $\Ls$.  Since $\cal P(\P)$ is a fundamental region for $\Lc$, by Lemma \ref{lemma:KeyTechnicalLemma}, it can be used to uniquely index the cosets of $\Lc/\Ls$. \hfill $\blacksquare$

Note that $\Gc$ is not explicitly needed, since the step $\x = \Gc \v b$ is equivalent to $\Hc \x = \v b$.  Since $\Hc$ is also triangular, $\Hc \x = \b b$ can be regarded as a linear system of equations easily solved because of the triangular structure.

\subsection{Indexing Triangular Matrix Lattices}

This subsection gives a systematic procedure for indexing of triangular matrix lattices.  The indexing operation is the inverse of encoding, that is, mapping the cosets of $\Lc/\Ls$ represented by $\x$ to information vectors $\b b$ and is denoted $\index(\x) = \b b$.  

The lattice codeword $\x = \enc(\b b)$ is:
\begin{align}
\x &= \Gc \b b - Q_{\Ls}(\Gc \b b) .
\end{align}
To find the inverse operation $\b b = \index(\x)$, observe the following.  Let $\wt b = \Hc \x$, so that:
\begin{align}
\wt b &= \b b - \Hc Q_{\Ls}(\Gc \b b) 
\end{align}
Let $\b c \in \bbzn$ satisfy $Q_{\Ls}(\Gc \b b) = \Gs \b c$ so that:
\begin{align}
\wt b &= \b b - \Hc \Gs \b c. \label{eqn:indexkey}
\end{align}
Using the triangular structure of $\Hc\Gs$, these equations are solved sequentially for $i=1$, then $i=2, \ldots, n$.  For convenience let $\Delta = \Hc \Gs$ which has entry  $\delta_{ij}$ in row $i$, column $j$: 
\begin{align}
\delta_{ij} &= \sum_{\ell=1}^n h_{i\ell} g_{\ell j},
\end{align}
and observe that $\delta_{ii} = M_i$.  

The first line of \eqref{eqn:indexkey} is
\begin{align}
\tilde b_1 &= b_1 - M_1 c_1,
\end{align}
which has solution $b_1$ and $c_1$ given by:
\begin{align}
b_1 &= \tilde b_1 \bmod M_1, \\
c_1 &= \frac{b_1 - \tilde b_1}{M_1}. 
\end{align}
And for following lines $k = 2, \ldots, n$:
\begin{align}
\widetilde b_k &= b_k - \big( \sum_{i=1}^{k-1} \delta_{ki} c_i \big) - M_k c_k 
\end{align}
which has solution $b_k$ and $c_k$ given by:
\begin{align}
b_k &= \widetilde b_k + \big( \sum_{i=1}^{k-1} \delta_{ki} c_i \big) \bmod M_k, \\
c_k &= \frac{b_k - \tilde b_k - \big( \sum_{i=1}^{k-1} \delta_{ki} c_i \big) }{M_k}.  
\end{align}

\subsection{Shaping Construction D Lattices}

Construction D-based lattices are lattices formed from nested linear binary codes.  Construction D lattices are appealing because they can have good coding gain and are constructed using familiar binary codes such as LDPC codes \cite{Sadeghi-it06}, turbo codes \cite{Sakzad-arxiv11} and BCH codes \cite{Conway-1999}.  The following example uses polar lattices \cite{Yan-isit13}, but the principles can be applied generally.

In this example, the shaping lattice $\Ls$ is a convolutional code lattice, obtained by applying Construction A to a convolutional code; see \cite{Erez-it05*3} and \cite{Molu-itw15}.  Convolutional code lattices have increasingly good shaping gain as the number of trellis states increases \cite{Forney-it92}. The complexity is reasonable, since quantization can be implemented with the Viterbi algorithm.  Since a convolutional code operates on a stream of data of arbitrary length, the trellis code can be terminated so that the resulting shaping lattice has the same dimension as the coding lattice.  

\newcommand{\code}{\cal D}

\begin{example}
\label{example:ConstructionD}
For the coding lattice, two-level the nested binary codes are $\code_1$ and $\code_2$ that satisfy $\code_2 \subseteq \code_1$.  By embedding binary numbers directly in the real space, this Construction D lattice can be expressed as $\Lc = \code_2 + 2 \code_1 + 4 \bbzn$.  For a binary polar code of length $n=8$, let $\P \otimes \P \otimes \P$ be the 8-by-8 Kronecker product matrix of $\P = [1 \ 0 \ ; 1 \ 1]$.  The generator matrix for $\code_1$ consists of columns 1, 2, 3 and 5 from the Kronecker product; the generator matrix for $\code_2$ consists of column 1 from the Kronecker product.  Applying Construction D, a generator matrix for this lattice is:
\begin{align}
\Gc &=
\begin{bmatrix}
1 & 0 & 0 & 0 & 0 & 0 & 0 & 0 \\ 
1 & 2 & 0 & 0 & 0 & 0 & 0 & 0 \\ 
1 & 0 & 2 & 0 & 0 & 0 & 0 & 0 \\ 
1 & 0 & 0 & 2 & 0 & 0 & 0 & 0 \\ 
1 & 2 & 2 & 0 & 4 & 0 & 0 & 0 \\ 
1 & 2 & 0 & 2 & 0 & 4 & 0 & 0 \\ 
1 & 0 & 2 & 2 & 0 & 0 & 4 & 0 \\ 
1 & 2 & 2 & 2 & 0 & 0 & 0 & 4  
\end{bmatrix}.
\end{align}
The interchange of rows 4 and 5 of the Kronecker product matrix provides the triangular form\footnote{Note that $\C_1$ is an
extended (8,4) Hamming code and $\C_2$ is the (8,1) repeat code. This lattice has coding gain of 0.7525 dB (1.1892), which is less than the coding gain of 3 dB (2) of the $E_8$ lattice.}.  The following is a check matrix for this lattice, 
\begin{align}
\Hc = \begin{bmatrix}
1 & 0 & 0 & 0 & 0 & 0 & 0 & 0 \\ 
\oh & \oh & 0 & 0 & 0 & 0 & 0 & 0 \\ 
\oh & 0 & \oh & 0 & 0 & 0 & 0 & 0 \\ 
\oh & 0 & 0 & \oh & 0 & 0 & 0 & 0 \\ 
\oq & \oq & \oq & 0 & \oq & 0 & 0 & 0 \\ 
\oq & \oq & 0 & \oq & 0 & \oq & 0 & 0 \\ 
\oq & 0 & \oq & \oq & 0 & 0 & \oq & 0 \\ 
0 & \oq & \oq & \oq & 0 & 0 & 0 & \oq  
\end{bmatrix} \label{eqn:polarCheckMatrix}
\end{align}
It is not the inverse of $\Gc$, but $\Hc$ is a check matrix for $\Lc$.  Refer to \cite[Sec.~III]{Sadeghi-it06} for forming Construction D check matrices from code matrices. 
\end{example}

The shaping lattice will be a scaled version of the following convolutional code lattice. Consider a memory 1 convolutional code with generator polynomials $D+1$ and $1$, where $D$ is the delay operator. If this code is terminated to a block length of 8, then the block code representation generator is the binary matrix with three generator vectors:
\begin{align}
\begin{bmatrix}
1 & 1 & 0 & 1 & 0 & 0 & 0 & 0 \\ 
0 & 0 & 1 & 1 & 0 & 1 & 0 & 0 \\ 
0 & 0 & 0 & 0 & 1 & 1 & 0 & 1  
\end{bmatrix}.  \label{eqn:ccblock}
\end{align}
Form a lattice by applying this block code to Construction A, resulting in the following generator matrix:
\begin{align}
\G_1 &=
\begin{bmatrix}
1 & 0 & 0 & 0 & 0 & 0 & 0 & 0 \\ 
0 & 1 & 0 & 0 & 0 & 0 & 0 & 0 \\ 
0 & 0 & 1 & 0 & 0 & 0 & 0 & 0 \\ 
1 & 0 & 0 & 2 & 0 & 0 & 0 & 0 \\ 
1 & 1 & 0 & 0 & 2 & 0 & 0 & 0 \\ 
0 & 1 & 1 & 0 & 0 & 2 & 0 & 0 \\ 
0 & 0 & 0 & 0 & 0 & 0 & 2 & 0 \\ 
0 & 0 & 1 & 0 & 0 & 0 & 0 & 2  
\end{bmatrix}, \label{eqn:ccllatice}
\end{align}
where columns \eqref{eqn:ccblock} (corresponding to rows of \eqref{eqn:ccllatice}) are interchanged to put $\G_1$ in triangular form.  This lattice has a shaping gain of approximately 0.25 dB, but convolutional lattice codes may have a much higher shaping gain.

Obtain the shaping lattice $\Ls$ by scaling this lattice by some integer $K$, that is $\Gs = K \G_1$.  The sublattice condition is satisfied for $K=4,8,12, \ldots$. Following Lemma \ref{lemma:triangularLattice}, \ri is accomplished using values for $M_1, M_2, \ldots, M_8$ as:
\begin{align}
K,\ \frac K 2 ,\  \frac K 2 ,\  K,\  \frac K 2 ,\ \frac K 2 ,\ \frac K 2 ,\ \frac K 2 ,
\end{align}
respectively.  The product of the $M_i$ is $M = K^8 / 64$, so the code rate is $R = \log_2 (K) - \frac 3 4$, and code rates corresponding to $K=4,8,12, \ldots$ are $1.25, 2.25, 2.8350, \ldots$ bits per dimension.

\subsection{Shaping Construction A Lattices}

Construction A-based lattices are lattices formed from a single linear code.  Non-binary codes can provide better coding gain than binary codes. LDA lattices, based on non-binary LDPC codes, have good error-correction properties, can be constructed in high dimension, but they must be constructed over the ring of integers modulo $p$, where $p$ is an prime \cite{diPietro-itw12}.

For shaping, small-dimension lattices such as $D_n$, $E_8$, Barnes-Wall and Leech lattice have excellent shaping gain for their respective dimension, and efficient quantization algorithms.  The dimension of such lattices is not matched to high-dimension coding lattices.  This problem can be solved by taking the  Cartesian product, and the corresponding generator matrix $\Gs$ has a block-diagonal form. The following example uses a two-fold Cartesian product, but more generally a $\frac n m$-fold Cartesian product can be used, if the unexpanded shaping lattice has dimension $m$.

\begin{example}
\label{example:ConstructionA} Construction A forms lattices from a code constructed over the $p$-ary ring of integers modulo $p$.  The coding lattice is based on a code with parity check matrix of block length $n=8$ and $p=5$ given by the following six parity checks:
\begin{align}
\begin{bmatrix}
0 & 1 & 2 & 0 & 0 & 0 & 0 & 0 \\ 
0 & 0 & 0 & 1 & 0 & 0 & 0 & 0 \\ 
4 & 0 & 0 & 0 & 4 & 0 & 0 & 0 \\ 
0 & 0 & 4 & 0 & 0 & 3 & 0 & 0 \\ 
4 & 0 & 0 & 3 & 0 & 0 & 2 & 0 \\ 
0 & 3 & 0 & 0 & 2 & 0 & 0 & 1  
\end{bmatrix}. 
\end{align}
This code is based on a modified array code \cite{Eleftheriou-icc02}, which has a triangular portion which is characteristic of some LDPC codes used in practice, it also means the matrix is also full-rank.  Under Construction A, the resulting lattice $\Lc$ has check matrix $\Hc$ given by:
\begin{align}
\Hc = \begin{bmatrix}
1 & 0 & 0 & 0 & 0 & 0 & 0 & 0 \\ 
0 & 1 & 0 & 0 & 0 & 0 & 0 & 0  \\
0 & \frac 1 5 & \frac 2 5 & 0 & 0 & 0 & 0 & 0 \\ 
0 & 0 & 0 & \frac 1 5 & 0 & 0 & 0 & 0 \\ 
\frac 4 5 & 0 & 0 & 0 & \frac 4 5 & 0 & 0 & 0 \\ 
0 & 0 & \frac 4 5 & 0 & 0 & \frac 3 5 & 0 & 0 \\ 
\frac 4 5 & 0 & 0 & \frac 3 5 & 0 & 0 & \frac 2 5 & 0 \\ 
0 & \frac 3 5 & 0 & 0 & \frac 2 5 & 0 & 0 & \frac 1 5  
\end{bmatrix} .
\end{align}
Refer to \cite[p.~33]{Zamir-2014} for forming Construction A matrices from code matrices. 
\end{example}

For shaping, a scaled version of the $D_4$ lattice is used, which has 0.37 dB shaping gain.  A Cartesian product of this lattice is used so the dimensions match. This Cartesian product lattice has the same shaping gain as the original lattice. The smallest scaling that satisfies the sublattice condition is 5, and so the shaping lattice is $\Ls = 5D_4 \times 5D_4$ with generator matrix given by:
\begin{align}
\Gs = \begin{bmatrix*}[r]
5 & 0 & 0 & 0 & 0 & 0 & 0 & 0 \\ 
-5 & 5 & 0 & 0 & 0 & 0 & 0 & 0 \\ 
0 & -5 & 5 & 0 & 0 & 0 & 0 & 0 \\ 
0 & 0 & -5 & 10 & 0 & 0 & 0 & 0 \\ 
0 & 0 & 0 & 0 & 5 & 0 & 0 & 0 \\ 
0 & 0 & 0 & 0 & -5 & 5 & 0 & 0 \\ 
0 & 0 & 0 & 0 & 0 & -5 & 5 & 0 \\ 
0 & 0 & 0 & 0 & 0 & 0 & -5 & 10  
\end{bmatrix*}.  
\end{align}
Quantization in $\Ls$ is performed by applying the standard algorithm \cite{Conway-1999} to each $D_4$ independently.  This readily generalizes, and other shaping lattices  mentioned earlier, including the Leech lattice, also have triangular generator matrices.

These lattices satisfy the sublattice condition, and $V(\Lc) = 5^6 / 3 \cdot 2^4$  and $V(\Ls) = 4 \cdot 5^8$, so $M=4800$ points can be encoded. Following Lemma \ref{lemma:triangularLattice}, \ri is accomplished using the values for $M_1, \ldots, M_8$ as:
\begin{align}
  5, 5, 2, 2, 4, 3, 2 ,2
\end{align}
respectively.  The resulting code rate is $R \approx 1.53$ bits/dimension.

\section{Encoding and Indexing Full Matrix Lattices}
\label{sec:full}

This section describes encoding and indexing lattice codes when $\Gc$ and $\Gs$ are not necessarily triangular. Specifically, the given $\Gc$ cannot be scaled to form a \ptope which is a fundamental region of $\Ls$, and so cannot be used for \ri.  This section shows how to find a new basis $\Gc'$ for $\Lc$, and corresponding diagonal matrix $\M$, such that the \ptope $\cal P(\Gc' \M)$ is a fundamental region of $\Ls$, and thus can be used for \ri.

\subsection{Basis Change for Encoding Full Matrix Lattices}

This subsection describes the basis change procedure. The new basis $\Gc'$ is defined as scaled versions of $n-1$ basis vectors from $\Gs$.  The last basis vector is selected to satisfy the condition that $\Gc'$ will generate $\Lc$.  If such a $\Gc'$ exists, then $\Lc / \Ls$ has a \ri.

Recall $\b g_1, \b g_2, \ldots, \b g_n$ are the basis vectors, columns of $\Gs$, and $\Ls \subseteq \Lc$. Then $\b g_i$ is an element of $\Lc$, and $\Hc \b g_i$ is an integer vector.   Define $m_i$ as the greatest common denominator of all elements of a vector:
\begin{align}
m_i = \gcd( \Hc \b g_i), \label{eqn:shortervector}
\end{align}
for $i=1,2, \ldots, n$.  The scaled vector $\frac{\b g_i}{m_i}$ is also an element of $\Lc$, since $\Hc \frac{\b g_i}{m_i} $ is also an integer vector ($\frac 1 {m_i}$ is the smallest scaling such that $\frac{\b g_i}{m_i}$ is still an element of $\Lc$).  Thus $\frac{\b g_i}{m_i}$ is a candidate basis vector. The following set of vectors:
\begin{align}
\begin{bmatrix}
\frac{ \b g_1}{m_1} & \frac{\b g_2}{m_2} &  \cdots & \frac{\b g_{n}}{m_{n}}  \label{eqn:notbasis} 
\end{bmatrix},
\end{align}
are linearly independent and each is a member of $\Lc$. But in general, these do not form a basis for $\Lc$.

Assume that $\Lc$ has a basis $\Gc'$ of the following form, where one vector in column $t$ is replaced with an unknown column vector $\b q$:
\newcommand{\mys}{\ \ \ }
\begin{align}
\Gc' =
\begin{bmatrix}
\frac{ \b g_1}{m_1} \mys \cdots \mys   \frac{\b g_{t-1}}{m_{t-1}} \mys   \v q \mys   \frac{\b g_{t+1}}{m_{t+1}} \mys \cdots \mys  \frac{\b g_{n}}{m_{n}} \label{eqn:modifiedbasis}
\end{bmatrix} .
\end{align}
This assumption may not always hold, but for a variety of cases, it was found to hold. A concrete method to search for $\v q$ is described, that is, if such a $\b q$ can be found, then $\Gc'$ is a basis for $\Lc$.

A basis transformation for $\Lc$ from $\Gc$ to $\Gc'$ is given by:
\begin{align}
\Gc' &= \Gc \W \oor \\
\Hc \Gc' &= \W
\end{align}
where $\W$ is a unimodular matrix, that is, it has integer entries and $| \det (\W)| = 1$.   The vector $\b q$ is selected to satisfy the condition that $\W$ is unimodular.  Write $\W$ as follows, for example, if $t=n$:
\begin{align}
\W &=
\begin{bmatrix}
w_{1,1} & w_{1,2} & \cdots & w_{1,n-1} & r_{1} \\
w_{2,1} & w_{2,2} & \cdots & w_{2,n-1} & r_{2} \\
\vdots & \vdots & & \vdots & \vdots \\
w_{n,1} & w_{n,2} & \cdots & w_{n,n-1} & r_{n} \\
\end{bmatrix}.
\end{align}
The integers $w_{i,j}$ in all columns except column $t$ are linearly dependent and are readily found.  The integers $r$ in column $t$ are selected to satisfy $|\det (\W) | = 1$, for a positive determinant that is:
\begin{align}
\det (\W) = \sum_{i=1}^n (-1)^{i+t} r_i \det (\W^{(i,t)}) = 1 \label{eqn:diophantine}
\end{align}
where $\W^{(i,t)}$ is the $n-1 \times n-1$ submatrix of $\W$ with row $i$ and column $t$ removed.  So that $\W$ is unimodular, we seek a solution to $ \det (\W) = 1$ where the variables $r_i$ are integers.  This is a linear diophantine equation in the variables $r_1, r_2, \ldots, r_n$. 

There may be multiple solutions, or there may be no solution.  If all coefficients of \eqref{eqn:diophantine} are even, then there is no solution.  If any two pairs of coefficients are relatively prime, then a solution exists by applying the extended Euclidean algorithm to those two coefficients, and setting other $r_i =0$. 

Given $\Gs$ for $\Ls$ and $\Gc$ for $\Lc$, the following lemma gives a condition on the existence of a \ri:
\begin{lemma}
Assume a solution $r_1, \ldots, r_n$ to $|\det (\W)| = 1$ exists.  Then a \ri can be obtained using the modified basis $\Gc'$ given by \eqref{eqn:modifiedbasis} with
\begin{align}
\v q = \Gc \cdot 
\begin{bmatrix}
r_1 \\ \vdots \\ r_n 
\end{bmatrix},
\end{align}
and 
\begin{align}
 M_i =
\begin{cases}
\frac{M}{\prod_{i=1}^{n\setminus t} m_i} & \iif i = t\\
m_i & \otherwise 
\end{cases},
\end{align}
where $m_i$ is given by \eqref{eqn:shortervector} and $M = \det(\Gs) / \det(\Gc)$.  
\end{lemma}

\emph{Proof} Consider the matrix $\P$:
\begin{align}
\P &= \Gc' \M.
\end{align}
Except for $i=t$, each column $i$ of $\P$ is $\v g_i$, a generator vector of $\Ls$.  Since $\W$ is a basis transformation, $\det(\Gc) = \det(\Gc')$. By Lemma \ref{lemma:squareFundamental}, $\cal P(\P)$ is a fundamental region for $\Ls$.  Since $\cal P(\P)$ is a fundamental region for $\Lc$, by Lemma \ref{lemma:KeyTechnicalLemma} this $\Gc'$ and $\M$ can be used to uniquely index the cosets of $\Lc/\Ls$ and form a \ri. \hfill $\blacksquare$

\subsection{Indexing Full Matrix Lattices}

This subsection gives a systematic procedure for indexing of full matrix lattices.  The indexing operation is the inverse of encoding, that is, mapping the cosets of $\Lc/\Ls$ represented by $\x$ to information vectors $\b b$ and is denoted $\index(\x) = \b b$. 

The lattice codeword $\x = \enc(\b b)$ is:
\begin{align}
\x &= \Gc' \b b - Q_{\Ls}(\Gc' \b b). 
\end{align}
To find the inverse operation $\b b = \index(\x)$, observe the following.  Let $\Hc' = (\Gc' )^{-1}$ and let $\wt b = \Hc' \x$ so that:
\begin{align}
\wt b = \b b - \Hc' \Gs \b c, \label{eqn:sqaureIndex}
\end{align}
where $\b c \in \bbzn$ satsifies $Q_{\Ls}(\Gc' \b b) = \Gs \b c$.  The matrix $\Hc' \Gs$ has the form:
\begin{align}
\begin{bmatrix*}
M_1 & 0  & \cdots & u_{1}  & \cdots & 0\\
0 & M_2  & \cdots & u_{2} & \cdots & 0 \\
\vdots & \vdots  &  &  \vdots &  & \vdots \\
0  & 0 &                      & M_{t} &  &0   \\
\vdots &\vdots &   & \vdots &  & \vdots \\
0 & 0 & \cdots & u_n & \cdots & M_n
\end{bmatrix*},
\end{align}
that is, it is $\diag(M_1, \ldots, M_n)$, with the change that column $t$ is a vector $\b u = \Hc' \b g_t$, with  $u_t = M_t$, and the other $u_i$ are non-zero in general.

Line $i$ of \eqref{eqn:sqaureIndex} is:
\begin{align}
\tilde b_i &=
\begin{cases}
 b_i + M_i c_i & i = t \\
 b_i + M_i c_i +u_i c_t & i \neq t
\end{cases} 
\end{align}
Then, the indexing procedure $\b b = \index(\x)$ is:
\begin{enumerate}
\item Find $\v u = \Hc' \v g_t$, where $\v g_t$ is column $t$ of $\Gs$.
\item Find $\wt b = \Gc' \x$.
\item For only $t$: $b_t = \tilde b_t \bmod M_t$, and $c_t = \frac{\tilde b_t - b_t}{M_t}$.
\item  For $i=1, \ldots, n$ except $t$: $b_i = (\tilde b_i - u_i c_t) \bmod M_i$
\end{enumerate}

\subsection{Example \ref{example:Full3} and \ref{example:Full8}: Encoding Using Full Matrix Lattices}

Two examples are given to illustrate encoding of full matrix lattices.  Example \ref{example:Full3} shows the mechanics of the encoding using $n=3$.  Full matrix lattice encoding was motivated by LDLC lattices, and Example \ref{example:Full8} illustrates encoding using an $n=8$ full matrix LDLC lattices.  
\begin{example}
\label{example:Full3} 
Consider the coding lattice $\Lc$ with $\Hc$ given by:
\begin{align}
\Hc = \begin{bmatrix*}[r]
1 & \oh & \oq \\ 
-\oq & 1 & 0 \\ 
0 & -\oq & 1  
\end{bmatrix*}  
\end{align}
and a shaping lattice $4 D_3$ given by:
\begin{align}
\Gs = 
\begin{bmatrix}
4 & 0 & 0 \\ 
-4 & 4 & 0 \\ 
0 & -4 & 8  
\end{bmatrix}  
\end{align}
which has shaping gain of $0.25$ dB.  The code rate is $R = \frac 1 3 \log_2\big( 128 / \frac{64}{73}  \big) \approx 2.40$ bits per dimension.

Using \eqref{eqn:shortervector}, $(m_1, m_2, m_3) = (1,1,2)$ and so following scaled versions of $\Gs$ are candidate basis vectors for $\Lc$:
\begin{align}
\begin{bmatrix}
4 & 0 & 0 \\ 
-4 & 4 & 0 \\ 
0 & -4 & 4  
\end{bmatrix}  
\end{align}
Choosing column $t=1$, the $\W$ matrix is:
\begin{align}
\W = \begin{bmatrix}
r_1 & 1 & 1 \\ 
r_2 & 4 & 0 \\ 
r_3 & -5 & 4  
\end{bmatrix},
\end{align}
and $\det(\W) = 1$ leads to the diophantine equation:
\begin{align}
16 r_1 - 9 r_2  - 4 r_1 = 1, 
\end{align}
which has a solution $(r_1,r_2,r_3) = (4,7,0)$.   The resulting modified check matrix is:
\begin{align}
\Hc' = (\Gc')^{-1} = 
\begin{bmatrix}
73/4 & 0 & 0 \\ 
-32 & 1/4 & 0 \\ 
-40 & 1/4 & 1/4  
\end{bmatrix}  
\end{align}
and the encoding range is $(M_1,M_2, M_3) = (73,1,2)$.
\end{example}

The next example uses an $n=8$ matrix that shows the structure of the LDLC check matrix. The check matrix $\Hc$ is specified by design, where there is a dominant 1 entry in each row and each column.  Other elements are selected so the matrix is sparse, with constant row and column weight, and random sign changes.

\begin{example}
\label{example:Full8}
This example chooses the non-dominant elements of $\Hc$ to be $\pm \oh, \pm \oq$ for:
 \begin{align*}
\Hc = \begin{bmatrix*}[r]
1 & 0 & 0 & \oh & 0 & 0 & -\oq & 0 \\ 
\oq & 1 & 0 & 0 & 0 & 0 & 0 & -\oh \\ 
0 & \oh & 1 & 0 & 0 & 0 & 0 & \oq \\ 
0 & 0 & \oq & 1 & 0 & 0 & -\oh & 0 \\ 
\oh & 0 & 0 & 0 & 1 & -\oq & 0 & 0 \\ 
0 & 0 & \oh & \oq & 0 & 1 & 0 & 0 \\ 
0 & -\oq & 0 & 0 & \oh & 0 & 1 & 0 \\ 
0 & 0 & 0 & 0 & -\oq & \oh & 0 & 1  
\end{bmatrix*} 
\end{align*}
which will satisfy the sublattice condition for the shaping lattice, which is $8 E_8$:
\begin{align*}
\Gs = \begin{bmatrix*}[r]
4 & 0 & 0 & 0 & 0 & 0 & 0 & 0 \\ 
4 & 8 & 0 & 0 & 0 & 0 & 0 & 0 \\ 
4 & -8 & 8 & 0 & 0 & 0 & 0 & 0 \\ 
4 & 0 & -8 & 8 & 0 & 0 & 0 & 0 \\ 
4 & 0 & 0 & -8 & 8 & 0 & 0 & 0 \\ 
4 & 0 & 0 & 0 & -8 & 8 & 0 & 0 \\ 
4 & 0 & 0 & 0 & 0 & -8 & 8 & 0 \\ 
4 & 0 & 0 & 0 & 0 & 0 & -8 & 16  
\end{bmatrix*}.
\end{align*}
The resulting lattice code has rate $R \approx 3.01$ bits per dimension.

The eight subdeterminants which form the coefficients in \eqref{eqn:diophantine}, in order, are: 
\begin{align*}
 -54018, 15597,  7778 , 45151, -3072, -3695, 9071, 5854.
\end{align*}
There are various approaches to finding a solution, but one is to note that 54018 and 45151 are relatively prime, and by Bezout's identity, choosing $r_1 = 15327$ and $r_4 = 18337$ and all other $r_i = 0$, is a solution.  Rectangular encoding is accomplished using values for $M_1,M_2, \ldots, M_8$ as:
\begin{align*}
1,2, 2, 137746, 2, 2, 2, 4.
\end{align*}
\end{example}
A consequence of choosing $t=4$ as the target for replacement is that most of the information integers are encoded into this position. Even though the check matrix of the modified basis, $\Hc' = (\Gc')^{-1}$ is not suitable for belief-propagation decoding, this is of no consequence since the lattice $\Lc$ itself is not changed, the original check matrix $\Hc$ may be used by the decoding algorithm. Since the encoding described in this paper deals with how information is mapped to lattice points, the structure of the lattice code is not modified, and the mapping does not affect the probability of error due to the decoder chooses a lattice point which is different from the transmitted lattice point.

\subsection{Cyclic Groups}

Under \ri, the lattice code $\C$ is a cyclic group, in the following cases.  Recall that $M = \prod_{i=1}^n M_i$.  The lattice code $\C$ forms a cyclic group if the \ri has, for some $k$, $M_k = M$ and $M_i = 1$ for $i \neq k$.  In this case, the generator vector $\b v_k$ in column $k$ of the generator matrix $\Gc$ (or $\Gc'$ if using a modified basis) can generate the entire code $\C$.  Define $\b b'$ as:
\begin{align}
\b b' = \begin{bmatrix}
0 & \cdots & 0 & b_k & 0 \cdots & 0
\end{bmatrix}\tr,
\end{align}
that is only position $k$ has a non-zero element,  apply this $\b b'$ to the encoding \eqref{eqn:encode}. Then the operation:
\begin{align}
\x =  b'_k \b v_k  - Q_{\Ls}(b'_k \b v_k )
\end{align}
with $b'_k \in \{0,1, \ldots, M-1\}$, will generate the entire group $\C$.  Thus, $\C$ is a cyclic group, with generator element $\b v_k$. 

Referring to Example \ref{fig:example1}, a \ri exists for $(M_1, M_2) = (1,36)$, which satisfies the condition to form a cyclic group, using $k=2$.  Thus, the element $[ \frac 2 9, \frac 8 9]$ is a generator for this cyclic group.  This is one of 12 generator elements for the cyclic group.




\section{Comments on Group Homomorphism}
\label{sec:homo}

\subsection{Homomorphism Existence Condition}

The group properties of lattice codes are important for compute-and-forward techniques for physical-layer network coding \cite{Nazer-it11}.  Feng, Silva and Kschischang defined a linear labeling of $\Lc$ which possesses well-defined group linearity properties.  Without making  assumptions about the underlying lattice code structure, they gave conditions on the existence of a homomorphism \cite{Feng-it13}.  In this section, a condition on a group homomorphism for lattice codes which does consider the structure, namely $\Gs$ and $\Hc$, is given.  Such a group homomorphism is potentially useful for compute-and-forward relaying.

Each element $b_i \in \{0,1, \ldots, M_{i-1}\}$ is regarded as an element from the ring of integers modulo $M_i$, $\bbz_{M_i} = \bbz / M_i \bbz$. The vector $\b b$ is from a group written as $\bbzstar$: 
\begin{align}
\bbzstar = \bbz_{M_1} \times  \bbz_{M_2} \times \cdots \times \bbz_{M_n}. 
\end{align}
Addition of two elements $\b b_1, \b b_2 \in \bbzstar$, denoted $\b b_1 \boxplus \b b_2$ is component-wise addition in $\bbzstar$.  

A group homomorphism between $\bbzstar$ with operation $\boxplus$ and the group $\C$ with operation $\oplus$ is desirable:
\begin{align}
\enc(\v b_1 \boxplus \v b_2) &= \enc(\v b_1) \oplus \enc(\v b_2).  \label{eqn:isomor1}
 \end{align}
This is the natural homomorphism in the sense it uses the group operations and mapping already given in this paper. The cardinality $|\C|$ and $|\bbzstar|$ of the two groups is equal, and there is a bijection between $\C$ and $\bbzstar$ under the \ri, for the lattice codes discussed in this paper.  The bijection is a homomorphism under a condition given by the following lemma.
\begin{lemma}
\label{lemma:isomorphism}
If all elements of row $i$ of $\Hc \Gs$ are divisible by $M_i$, for all $i=1,2, \ldots, n$, then
the group $\bbzstar$ with operation $\boxplus$ is homomorphic with the group $\C$ with operation $\varoplus$, under the function:
\begin{align}
\enc : \bbzstar \to \C.
\end{align}
\end{lemma}

\emph{Proof} Observe that $\enc(\v b_1 \boxplus \v b_2)$ and $\enc(\v b_1) \oplus \enc(\v b_2)$ are both in $\C$; if these two are in the same coset, then they are equal, since $\C$ is the coset leaders.  Consider that $Q_{\Ls}(\x)$ can be expressed as $\Gs \b c$, for some integer vector $\b c$. Applying this to the definition of encoding \eqref{eqn:encode}, \eqref{eqn:isomor1} becomes:
\begin{align*}
\Gc (\b b_1 + \b b_2 \bmod \M) + \Gs \b c_1 &= \Gc \b b_1 + \Gc \b b_2 + \Gs \b c_2,
\end{align*}
and after multiplication by $\Hc$:
\begin{align}
(\b b_1 + \b b_2) \bmod \M &= \b b_1 + \b b_2 + \Hc \Gs (\b c_2 - \b c_1). \label{eqn:stepOfProof}
\end{align}
Let $\b d_i$ be row $i$ of $\Hc \Gs$.  Then row $i$ of \eqref{eqn:stepOfProof} is:
\begin{align}
(b_{1,i} + b_{2,i}) \bmod M_i &= b_{1,i} + b_{2,i} + \b d_i (\b c_2 - \b c_1).
\end{align}
This equality will hold for all $b_{1,i}$ and $b_{2,i}$ if all elements of $\b d_i$ are divisible by $M_i$, which is the condition in the lemma.  Note that $\b c_2 - \b c_1$ satisfies the equality, since $\enc(\v b_1 \boxplus \v b_2)$ and $\enc(\v b_1) \oplus \enc(\v b_2)$ are both coset leaders, that is $\b c_2 - \b c_1$ is unique.  \hfill $\blacksquare$

If a homomorphism exists for $\enc$, then the inverse mapping $\index$ must preserve the group properties, forming an isomorphism, as well, but the comments here concentrate on the homomorphism.  

For self-similar lattice codes the homomorphism holds.  Let $\Gs = K \Gc$ for some integer $K \in \bbz$, and $M_i = K$ for all $i$.  Then, $\Hc \Gs$ is a diagonal matrix $K \I_n$, where $\I_n$ is the identity matrix.  This $K \I_n$ satisfies the condition of Lemma \ref{lemma:isomorphism}, that is, each element of row $i$ is divisible by $K$, for all rows $i$.

For hypercube shaping, the homomorphism \eqref{eqn:isomor1} holds when $\Gc$ is triangular.  This is discussed in the following subsection.

These two shaping approaches which satisfy Lemma \ref{lemma:isomorphism} have already been used in finite-length coding.  For example, Feng et al.~used hypercube shaping \cite{Feng-it13} for physical-layer network coding.  Sakzad, Viterbo, Bourtros and Hong used self-similar lattice shaping for compute-and-forward relaying \cite{Sakzad-isit14}.

However, the condition in Lemma \ref{lemma:isomorphism} is not satisfied for Examples 1--5, and it is not immediately obvious if homomorphisms exist for either the triangular matrix lattice or the full matrix lattice encodings described in this paper.  Nonetheless, Lemma \ref{lemma:isomorphism} expresses a design rule which may aid finding lattices $\Lc$ and $\Ls$ which not only are easily encodable, but also possess a homomorphic property. 
 
\subsection{Hypercube Shaping}

While this paper concentrates on shaping lattices using the Voronoi region, hypercube shaping fits the model developed in this paper.  The homomorphism \eqref{eqn:isomor1} under hypercube shaping can be shown.  These results hold when $\Gc$ is triangular. A hypercube $\cal H$ with side length $K$ is:
\begin{align}
\cal H &= \{ \x \in \bbrn \mid -K/2 \leq x_i < K/2 \},
\end{align}
and the code is $\cal C = \cal H \cap \Lc$. Let $M_i = K / v_{ii}$, where $v_{ii}$ are the diagonal elements of $\Gc$; note that $v_{ii}$ must satisfy the condition $K / v_{ii}$ is an integer. Define the shaping lattice for hypercubic shaping as $\Gs = \Gc \M$ (one might expect $\Gs = K \bbzn$, but this choice does not lead to a homomorphism).  

This $\cal H$ can be shown to be a fundamental region of the lattice $\Ls$.  This $\cal H$ (and not the Voronoi region of $\Ls$) is used to select the coset representatives. Quantization in $\Ls$ with respect to $\cal H$ is well defined: for any $\y \in \bbrn$, $\x = Q_{\cal H}(\y)$ means $\y \in \cal H + \x$, where $\x \in \Ls$.  Encoding to the hypercube codebook $\cal H \cap \Lc$ is obtained by:
\begin{align}
\enc(\b b) &= \Gc \b b - Q_{\cal H}(\Gc \b b),
\end{align}
for $b_i \in \{0,1, \ldots, M_i-1\}$ and $i = 1,2, \ldots, n$. 

Lemma \ref{lemma:isomorphism} and its proof applies for hypercube shaping as well since $Q_{\cal H}(\Gc \b b) = \Gs \b c$ for some $\b c \in \bbzn$.  Then, $\Hc \Gs = \M$ is a diagonal matrix where row $i$ is divisible by $M_i$.   Since the condition of Lemma \ref{lemma:isomorphism} are satisfied, the group homomorphism exists for hypercube-shaped lattice codes.

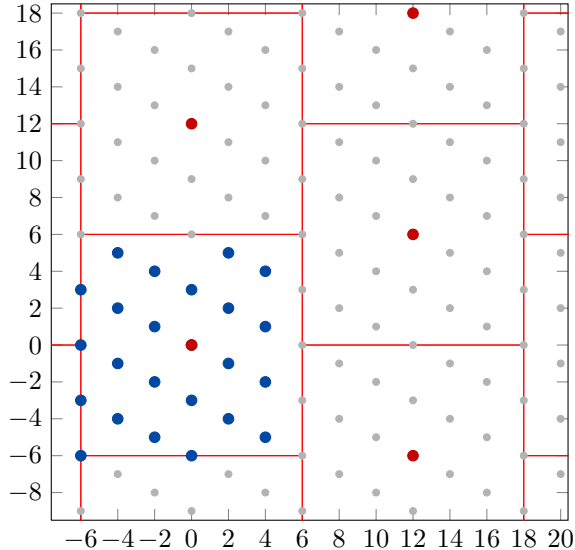
\begin{figure}[t]
\begin{center}
%
%
\definecolor{mycolor1}{rgb}{0.70196,0.70196,0.70196}%
\definecolor{mycolor2}{rgb}{0.00000,0.28235,0.63137}%
\definecolor{mycolor3}{rgb}{0.76863,0.00784,0.00784}%
\begin{tikzpicture}[%
scale=1
]

\begin{axis}[%
width=6.87cm,
height=6.87cm,
at={(0cm,0cm)},
scale only axis,
xmin=-7.6,
xmax=20.4,
xtick={-100,  -98,  -96,  -94,  -92,  -90,  -88,  -86,  -84,  -82,  -80,  -78,  -76,  -74,  -72,  -70,  -68,  -66,  -64,  -62,  -60,  -58,  -56,  -54,  -52,  -50,  -48,  -46,  -44,  -42,  -40,  -38,  -36,  -34,  -32,  -30,  -28,  -26,  -24,  -22,  -20,  -18,  -16,  -14,  -12,  -10,   -8,   -6,   -4,   -2,    0,    2,    4,    6,    8,   10,   12,   14,   16,   18,   20,   22,   24,   26,   28,   30,   32,   34,   36,   38,   40,   42,   44,   46,   48,   50,   52,   54,   56,   58,   60,   62,   64,   66,   68,   70,   72,   74,   76,   78,   80,   82,   84,   86,   88,   90,   92,   94,   96,   98,  100},
ymin=-9.5,
ymax=18.5,
ytick={-100,  -98,  -96,  -94,  -92,  -90,  -88,  -86,  -84,  -82,  -80,  -78,  -76,  -74,  -72,  -70,  -68,  -66,  -64,  -62,  -60,  -58,  -56,  -54,  -52,  -50,  -48,  -46,  -44,  -42,  -40,  -38,  -36,  -34,  -32,  -30,  -28,  -26,  -24,  -22,  -20,  -18,  -16,  -14,  -12,  -10,   -8,   -6,   -4,   -2,    0,    2,    4,    6,    8,   10,   12,   14,   16,   18,   20,   22,   24,   26,   28,   30,   32,   34,   36,   38,   40,   42,   44,   46,   48,   50,   52,   54,   56,   58,   60,   62,   64,   66,   68,   70,   72,   74,   76,   78,   80,   82,   84,   86,   88,   90,   92,   94,   96,   98,  100},
axis background/.style={fill=white}
]
\addplot [color=mycolor1,mark size=1.3pt,only marks,mark=*,mark options={solid},forget plot]
  table[row sep=crcr]{%
0	0\\
0	3\\
0	6\\
0	9\\
0	12\\
0	15\\
0	18\\
0	21\\
0	-9\\
0	-6\\
0	-3\\
2	-1\\
2	2\\
2	5\\
2	8\\
2	11\\
2	14\\
2	17\\
2	20\\
2	-10\\
2	-7\\
2	-4\\
4	-2\\
4	1\\
4	4\\
4	7\\
4	10\\
4	13\\
4	16\\
4	19\\
4	-11\\
4	-8\\
4	-5\\
6	-3\\
6	0\\
6	3\\
6	6\\
6	9\\
6	12\\
6	15\\
6	18\\
6	21\\
6	-9\\
6	-6\\
8	-4\\
8	-1\\
8	2\\
8	5\\
8	8\\
8	11\\
8	14\\
8	17\\
8	20\\
8	-10\\
8	-7\\
10	-5\\
10	-2\\
10	1\\
10	4\\
10	7\\
10	10\\
10	13\\
10	16\\
10	19\\
10	-11\\
10	-8\\
12	-6\\
12	-3\\
12	0\\
12	3\\
12	6\\
12	9\\
12	12\\
12	15\\
12	18\\
12	21\\
12	-9\\
14	-7\\
14	-4\\
14	-1\\
14	2\\
14	5\\
14	8\\
14	11\\
14	14\\
14	17\\
14	20\\
14	-10\\
16	-8\\
16	-5\\
16	-2\\
16	1\\
16	4\\
16	7\\
16	10\\
16	13\\
16	16\\
16	19\\
16	-11\\
18	-9\\
18	-6\\
18	-3\\
18	0\\
18	3\\
18	6\\
18	9\\
18	12\\
18	15\\
18	18\\
18	21\\
20	-10\\
20	-7\\
20	-4\\
20	-1\\
20	2\\
20	5\\
20	8\\
20	11\\
20	14\\
20	17\\
20	20\\
22	-11\\
22	-8\\
22	-5\\
22	-2\\
22	1\\
22	4\\
22	7\\
22	10\\
22	13\\
22	16\\
22	19\\
-10	5\\
-10	8\\
-10	11\\
-10	14\\
-10	17\\
-10	20\\
-10	-10\\
-10	-7\\
-10	-4\\
-10	-1\\
-10	2\\
-8	4\\
-8	7\\
-8	10\\
-8	13\\
-8	16\\
-8	19\\
-8	-11\\
-8	-8\\
-8	-5\\
-8	-2\\
-8	1\\
-6	3\\
-6	6\\
-6	9\\
-6	12\\
-6	15\\
-6	18\\
-6	21\\
-6	-9\\
-6	-6\\
-6	-3\\
-6	0\\
-4	2\\
-4	5\\
-4	8\\
-4	11\\
-4	14\\
-4	17\\
-4	20\\
-4	-10\\
-4	-7\\
-4	-4\\
-4	-1\\
-2	1\\
-2	4\\
-2	7\\
-2	10\\
-2	13\\
-2	16\\
-2	19\\
-2	-11\\
-2	-8\\
-2	-5\\
-2	-2\\
};
\addplot [color=mycolor2,mark size=2.0pt,only marks,mark=*,mark options={solid},forget plot]
  table[row sep=crcr]{%
0	0\\
0	3\\
0	-6\\
0	-3\\
2	-1\\
2	2\\
2	5\\
2	-4\\
4	-2\\
4	1\\
4	4\\
4	-5\\
-6	3\\
-6	-6\\
-6	-3\\
-6	0\\
-4	2\\
-4	5\\
-4	-4\\
-4	-1\\
-2	1\\
-2	4\\
-2	-5\\
-2	-2\\
};
\addplot [color=mycolor3,mark size=2.0pt,only marks,mark=*,mark options={solid},forget plot]
  table[row sep=crcr]{%
0	0\\
0	12\\
0	24\\
0	36\\
0	-12\\
12	-6\\
12	6\\
12	18\\
12	30\\
12	-18\\
24	-12\\
24	0\\
24	12\\
24	24\\
24	36\\
36	-18\\
36	-6\\
36	6\\
36	18\\
36	30\\
-12	6\\
-12	18\\
-12	30\\
-12	-18\\
-12	-6\\
};
\addplot [color=red,solid,forget plot]
  table[row sep=crcr]{%
-30	-18\\
-18	-18\\
-18	-6\\
-30	-6\\
-30	-18\\
};
\addplot [color=red,solid,forget plot]
  table[row sep=crcr]{%
-30	-6\\
-18	-6\\
-18	6\\
-30	6\\
-30	-6\\
};
\addplot [color=red,solid,forget plot]
  table[row sep=crcr]{%
-30	6\\
-18	6\\
-18	18\\
-30	18\\
-30	6\\
};
\addplot [color=red,solid,forget plot]
  table[row sep=crcr]{%
-30	18\\
-18	18\\
-18	30\\
-30	30\\
-30	18\\
};
\addplot [color=red,solid,forget plot]
  table[row sep=crcr]{%
-30	30\\
-18	30\\
-18	42\\
-30	42\\
-30	30\\
};
\addplot [color=red,solid,forget plot]
  table[row sep=crcr]{%
-30	42\\
-18	42\\
-18	54\\
-30	54\\
-30	42\\
};
\addplot [color=red,solid,forget plot]
  table[row sep=crcr]{%
-18	-24\\
-6	-24\\
-6	-12\\
-18	-12\\
-18	-24\\
};
\addplot [color=red,solid,forget plot]
  table[row sep=crcr]{%
-18	-12\\
-6	-12\\
-6	0\\
-18	0\\
-18	-12\\
};
\addplot [color=red,solid,forget plot]
  table[row sep=crcr]{%
-18	0\\
-6	0\\
-6	12\\
-18	12\\
-18	0\\
};
\addplot [color=red,solid,forget plot]
  table[row sep=crcr]{%
-18	12\\
-6	12\\
-6	24\\
-18	24\\
-18	12\\
};
\addplot [color=red,solid,forget plot]
  table[row sep=crcr]{%
-18	24\\
-6	24\\
-6	36\\
-18	36\\
-18	24\\
};
\addplot [color=red,solid,forget plot]
  table[row sep=crcr]{%
-18	36\\
-6	36\\
-6	48\\
-18	48\\
-18	36\\
};
\addplot [color=red,solid,forget plot]
  table[row sep=crcr]{%
-6	-30\\
6	-30\\
6	-18\\
-6	-18\\
-6	-30\\
};
\addplot [color=red,solid,forget plot]
  table[row sep=crcr]{%
-6	-18\\
6	-18\\
6	-6\\
-6	-6\\
-6	-18\\
};
\addplot [color=red,solid,forget plot]
  table[row sep=crcr]{%
-6	-6\\
6	-6\\
6	6\\
-6	6\\
-6	-6\\
};
\addplot [color=red,solid,forget plot]
  table[row sep=crcr]{%
-6	6\\
6	6\\
6	18\\
-6	18\\
-6	6\\
};
\addplot [color=red,solid,forget plot]
  table[row sep=crcr]{%
-6	18\\
6	18\\
6	30\\
-6	30\\
-6	18\\
};
\addplot [color=red,solid,forget plot]
  table[row sep=crcr]{%
-6	30\\
6	30\\
6	42\\
-6	42\\
-6	30\\
};
\addplot [color=red,solid,forget plot]
  table[row sep=crcr]{%
6	-36\\
18	-36\\
18	-24\\
6	-24\\
6	-36\\
};
\addplot [color=red,solid,forget plot]
  table[row sep=crcr]{%
6	-24\\
18	-24\\
18	-12\\
6	-12\\
6	-24\\
};
\addplot [color=red,solid,forget plot]
  table[row sep=crcr]{%
6	-12\\
18	-12\\
18	0\\
6	0\\
6	-12\\
};
\addplot [color=red,solid,forget plot]
  table[row sep=crcr]{%
6	0\\
18	0\\
18	12\\
6	12\\
6	0\\
};
\addplot [color=red,solid,forget plot]
  table[row sep=crcr]{%
6	12\\
18	12\\
18	24\\
6	24\\
6	12\\
};
\addplot [color=red,solid,forget plot]
  table[row sep=crcr]{%
6	24\\
18	24\\
18	36\\
6	36\\
6	24\\
};
\addplot [color=red,solid,forget plot]
  table[row sep=crcr]{%
18	-42\\
30	-42\\
30	-30\\
18	-30\\
18	-42\\
};
\addplot [color=red,solid,forget plot]
  table[row sep=crcr]{%
18	-30\\
30	-30\\
30	-18\\
18	-18\\
18	-30\\
};
\addplot [color=red,solid,forget plot]
  table[row sep=crcr]{%
18	-18\\
30	-18\\
30	-6\\
18	-6\\
18	-18\\
};
\addplot [color=red,solid,forget plot]
  table[row sep=crcr]{%
18	-6\\
30	-6\\
30	6\\
18	6\\
18	-6\\
};
\addplot [color=red,solid,forget plot]
  table[row sep=crcr]{%
18	6\\
30	6\\
30	18\\
18	18\\
18	6\\
};
\addplot [color=red,solid,forget plot]
  table[row sep=crcr]{%
18	18\\
30	18\\
30	30\\
18	30\\
18	18\\
};
\addplot [color=red,solid,forget plot]
  table[row sep=crcr]{%
30	-48\\
42	-48\\
42	-36\\
30	-36\\
30	-48\\
};
\addplot [color=red,solid,forget plot]
  table[row sep=crcr]{%
30	-36\\
42	-36\\
42	-24\\
30	-24\\
30	-36\\
};
\addplot [color=red,solid,forget plot]
  table[row sep=crcr]{%
30	-24\\
42	-24\\
42	-12\\
30	-12\\
30	-24\\
};
\addplot [color=red,solid,forget plot]
  table[row sep=crcr]{%
30	-12\\
42	-12\\
42	0\\
30	0\\
30	-12\\
};
\addplot [color=red,solid,forget plot]
  table[row sep=crcr]{%
30	0\\
42	0\\
42	12\\
30	12\\
30	0\\
};
\addplot [color=red,solid,forget plot]
  table[row sep=crcr]{%
30	12\\
42	12\\
42	24\\
30	24\\
30	12\\
};
\end{axis}
\end{tikzpicture}%
\caption{Hypercube shaping for Example 6, with the codebook $\C$ in blue. The hypercube with $K=12$ could be obtained as the zero-centered Voronoi region of $12 \bbz^2$, but this would not yield a homomorphism.  Instead, this figure shows the shaping lattice in red as $\Gc \M$, and the hypercube $\cal H$ is a fundamental region of this lattice.   \label{fig:example6} }
\end{center}
\end{figure}

It is interesting (and perhaps unexpected) that the shaping lattice is not $K \bbzn$, and this is illustrated with an example.  

\begin{example}
Let the coding lattice be given by:
\begin{align}
\Gc = 
\begin{bmatrix*}[r]
2 & 0 \\ 
-1 & 3  
\end{bmatrix*} 
\end{align}
and let $K = 12$.  The resulting hypercube code has $M= \frac{K^n}{\det(\Gc)} = 24$ codewords, and the scaling is $(M_1, M_2) = (6,4)$. \figref{fig:example6} illustrates the shaping lattice in red.  The lattices generated by $\Gc \M$ and $K \bbzn$ both have a zero-centered square as the shaping region.  The point of the figure is that rather than the regular structure of the $K \bbzn$ shaping lattice, it is the $\Gc \M$ lattice with square fundamental regions that provides the group homomorphism.
\end{example}

\section{Discussion}
\label{sec:discussion}

Shaped codes for the point-to-point AWGN communication channel can improve power efficiency by as much as 1.53 dB over hypercube constellations.  Self-similar lattices are not always the best choice, because of the complexity of the quantization algorithm needed to implement the modulo-lattice operation.  Alternatively, it is possible to use another lattice, one with an efficient quantization algorithm, to perform shaping instead.  While it is always possible to form such a lattice code, this paper's contribution is to deal with an important and practical aspect: mapping information to lattice codewords.

The best-case scenario is when the coding lattice has a generator matrix in triangular form.  If so, then any of the well-known lattices ($D_n$ to Leech lattice) or convolutional code lattices can be used for shaping, because these also have triangular generator matrices. Then efficient encoding is possible, as Lemma \ref{lemma:triangularLattice} showed.  However, finding any lattice generator matrix, particularly for Poltyrev-bound approaching coding lattices, is not always straightforward.  Furthermore, lattices with triangular generator matrices may not be good lattices.  For example, modified array codes used to construct LDPC codes with a triangular matrix can be used to obtain a lattice matrix in triangular form, but these codes are not especially good as the lattice dimension increases.  Similarly, triangular matrix LDLC lattices exist, but the triangular structure means certain elements are poorly protected from noise, and only ad hoc methods design methods have been used to overcome this problem, so far.

If lattice triangular generator matrices are not available, it nonetheless may still be possible to perform encoding using full matrices.  This requires finding a modified basis, which depends on the structure of the lattice and its generator matrix, which unlike the triangular matrix case, may not be as transparent.  As the lattice dimension grows, the magnitude of the integer range grows, particularly $M_t$ in column $t$, unless the coding lattice and shaping lattice are co-designed to avoid this problem.  In addition, this could possibly be employed as a component in another method, for example, encoding lattices described by block-wise triangular matrices.

The engineering benefit is that obtaining shaping gain of lattice codes appears to be feasible with reasonable complexity.  For wireless communication systems, this means an increase in spectral efficiency for point-to-point communications.  Reasonable complexity means that quantization operation in \eqref{eqn:encode} can be performed in a practical manner.  There appear to be two approaches to efficient lattice quantization.  One is to use one of the low-dimensional lattices such as $D_n$, $E_8$, Barnes-Wall ($n=16, 32, \ldots$) or Leech lattice ($n=24$), which have good shaping gain, and efficient quantization algorithms; Cartesian products of these lattices can be used to shape coding lattices of higher dimension, as was shown in Example \ref{example:ConstructionA}. The other approach is to use lattices with a trellis representation and quantization can be performed using the Viterbi algorithm.  Example \ref{example:ConstructionD} used a convolutional code lattice, and such a construction should be appealing since convolutional codes are already widely understood.

The results discussed thus far are applicable to lattices used for point-to-point communications.  If in addition, the lattice code is going to be used for physical layer network coding, such as compute-and-forward relaying, then a linear mapping between the information integers, and lattice code is needed. A necessary condition for a particular homomorphism to exist was given.  This condition is satisfied by self-similar lattices, and by hypercube-shaped lattices.  However, for more general lattice codes, including the examples in this paper, the condition is not readily satisfied.  Nonetheless, this condition can guide the design of lattice codes that simultaneously achieve good coding gain, efficient shaping algorithms, and homomorphisms for physical layer network coding.

\bibliographystyle{IEEEtrannourl}
\bibliography{/bibtex/abbrev,/bibtex/bitsbib,/bibtex/bits-noreview,/bibtex/bits-invited,/bibtex/bits-journal,/bibtex/bits-recent,/bibtex/bits-conf} 

\end{document}